\documentclass[12pt]{article}

 \usepackage{paralist} 
     \usepackage{placeins} 
     \usepackage{amsmath,amsthm,amssymb}
     \usepackage{epsfig}
     \usepackage{psfrag}
     \usepackage{tabls}
     \usepackage{supertabular}
     \usepackage[square,comma,numbers]{natbib}
     \usepackage[top=2.5cm, bottom=2.5cm, left=2.5cm, right=2.5cm]{geometry}
\usepackage{indentfirst} 
\usepackage{fancyheadings}
\usepackage{colortbl} 
\definecolor{niceyellow}{rgb}{0.98,0.92,0.73}
    \newcommand{\myfecha}{\today}

\begin{document}
\begin{center}
\Large{\bf On the teaching and learning of physics: A Criticism and 
           a Systemic Approach}
\end{center}
\begin{flushleft}
{\bf Sergio Rojas}\\
srojas@usb.ve \\
{Physics Department, Universidad Sim\'{o}n Bol\'{\i}var,
Ofic. 220, Apdo. 89000, Caracas 1080A, Venezuela.}
\end{flushleft}

{\bf Abstract}.
  The amount of published research in Physics Education Research (PER)
shows,
on one hand, an increasing interest in the design and development of
high performance
 physics teaching strategies, and, on the other hand, it
tries to  understand plausible
 ways on which the brain processes scientific information so that
scientific thinking skills could be taught more
effectively. As physics is a subject in which
mathematical and conceptual reasoning can not be
separated, instructors of
physics face the problem of finding suitable advise on the most
effective methods
of teaching physics (i.e. how much
time should be
spent on intuitive conceptual reasoning and how much time in developing
quantitative reasoning, and how to teach both in a unitary way).
In spite the
important
efforts made by the PER community, the published results
are overwhelming and
confusing for the physics instructors in the sense that the conclusions
that have arisen in those articles are in some instances controversial
and far from being
conclusive in pointing out a particular strategy to overcome the
afore mentioned problem.
Accordingly, based on the analysis of published PER work,  we'll be
arguing that
one of the major difficulties to overcome in the teaching of physics
could be associated to
the lack of
a consistent and coherent  methodological framework for teaching
which integrates both aspects,  conceptual and mathematical reasoning,
in a systemic
way of thinking. We will be presenting a set of plausible steps that
could be applied to tackle the aforementioned difficulty.

\bigskip
--More--(15
\setlength{\parindent}{0.75cm}

\section{Introduction}
  It is not difficult to find
  instructors of general physics courses anxiously perusing articles
  published by the Physics Education Research (PER) community, searching 
essentially for advice on how to approach the teaching of physics 
effectively. 
Despite the demand for actually useful pedagogy of physics,
PER has not produced 
so far any ultimately theory in this regards, and the great amount of published
research on the subject, in addition of being controversial\cite{Lea:1994,
Hammer:1996,Ehrlich:2002,Meltzer:2002,HoellwarthEtAll:2005,
Redish:2005,Ates:2007,Coletta:2008,Ates:2008}
might be overwhelming and even confusing to the physics instructors in
the sense that being physics intrinsically a quantitative based subject, 
much of the recent research 
favors an
overemphasis
on  qualitative (conceptual) physical aspects
\cite{MualemEylon:2007,HoellwarthEtAll:2005,SabellaRedish:2007, Walsh:2007},
while standard mathematical abilities, which are crucial for understanding
physical processes, are not stressed, or even taught, because, 
rephrasing a passage
from a recent editorial\cite{Klein:2007}, they
interfere with the students' emerging sense of physical insight.
Consequently, physics instructors 
face the problem of finding suitable advice on how to approach
the teaching of physics in the most efficient way and an answer to the
question of how much
time should be
spent on intuitive conceptual reasoning and how much time in developing
quantitative reasoning. Let's mention, in passing, that the
aforementioned  editorial\cite{Klein:2007} has risen an
interesting debate\cite{KleinDebate1:2007,KleinDebate2:2007,KleinDebate3:2007} 
regarding the benefits and 
shortcomings of
science education reform in the United States in relation to its influence
on the development of 
reasoning skills 
on the students, expressed in the way students use or apply the learned 
materials in their courses. 
The questions
raised in this debate are of interest to be taken into 
consideration by 
the PER community and/or curriculum developers, particularly
by those involved on science
education reform in developing countries.

On the other hand, physics instructors need also be mindful of the importance
of selecting the most appropriately functional
textbook, basically because 
 innovative
active-learning teaching schemes requires students to acquire basic and
fundamental knowledge through reading textbooks.
Correspondingly,
innovative teaching strategies should be designed to help
students
in processing their ever thicker and
heavier textbooks, which are laden with
physical and mathematical insights\cite{YapWong:2007,HoellwarthEtAll:2005,
CrouchMazur:2001, James:2006,CerbinKopp:2006}. Thus,
the panorama regarding the learning of physics is even more dramatic  
on the side of the students. For one reason,
in their struggle to fully participate in the process of learning,
at the moment of trying to find suitable learning materials that could help 
them to go beyond
classroom instruction (i.e. aiming to develop self-confidence on their
own through exercising their role of active-learners), students face the 
dilemma of deciding
which textbook could be helpful: perhaps a conceptual physics 
textbook (i.e. \cite{Hewitt:1993}),
the student could wonder; or may be a calculus based physics textbook
(i.e. \cite{HallidayResnickWalker:2000}); or
why not an algebra based physics textbook
(i.e. \cite{Giancoli:2004}); or what about a combination
of all of them? could finally the student ponder scrabbling 
in his/her
pocket/handbag to verify if the money in there could be enough to bring some
extra weight at home (for a good account of the drama of choosing a textbook 
see for instance \cite{Dake:2007} and references there in). 
For another, in a typical course work  
for students majoring in science and/or engineering  they usually
 must take more 
than one physics class. It could happen that
in one term his/her physics 
instructor may  emphasize
quantitative reasoning over conceptual analysis, and in another
term the respective instructor could rather accentuate conceptual
learning over quantitative analysis, likely causing confusion for students,
leading them to wonder which emphasis is correct.

Finally, it is not difficult to find published
results by the PER community on which it is shown directly
or indirectly the inability of students to express, interpret, and manipulate
physical results in mathematical terms. That is, students shows a clear
deficiency
in their training to exploit the mathematical solution of a problem
(which sometimes could be obtained mechanically or by rote procedures) to 
enhance their
knowledge regarding conceptual physics
\cite{Hammer:1996, Ehrlich:2002,Loverude:2003, SabellaRedish:2007,
RimoldiniSingh:2005,
      Meredith:2008, Rojas:2008}.
More important, the analysis of published  excerpts 
of 
student's responses to 
interviews
conducted by some researchers 
 to further
understand students' way of reasoning while solving physics problems,
shows that students   
 lack of a structured methodology for solving physics problems 
\cite{Hammer:1996, RimoldiniSingh:2005, SabellaRedish:2007,
Walsh:2007}. These findings
can not be surprising at all. In fact, none of the most commonly
recommended physics textbooks 
(i.e. \cite{HallidayResnickWalker:2000,TiplerMosca:2003,SerwayJewett:2003})
make use of a consistently and clear problem-solving
methodology when presenting the solution
of the textbook worked out illustrative examples 
\cite{tipler:note1}. Moreover, the
lack of a coherent problem-solving strategy can also be found in 
both the student and instructor
manual solutions that usually accompany textbooks. Generally, 
standard textbooks problem-solving strategies encourage the use of
a formula based scheme as compiled by the formulae
summary found at the end of each chapter 
of the text, and
this strategy seems to be spread out even in classroom teaching
\cite{Hamed:2008}.
Consequently, students 
merely imitates the way in which problems are handled in
the textbooks, 
which is also perhaps the same way in which
problems are solved by the instructor
in class.
For support of the previous assertion,
one needs only browse the Internet for introductory physics courses
and skim the solutions of problems posted by the course instructor.
Furthermore, from the aforementioned students interviews excerpts one 
can also appreciate
the lack of reasoning skills trying to associate or connect a way
to solving a problem with the solution of other previously similar
problem from another context (i.e. by using analogies). Again, the absence 
of this skill can not be
surprising at all because students are just mirroring the unrelatedness way 
on which 
commonly used physics textbooks present the themes (i. e. the use of analogies 
are not fostered)  
\cite{Schneider:2004, Donohoe:2008, tipler:note2}. 

\section{On the importance of a structured, systemic methodology to solve 
         physics problems}

To further motivate the subsequent discussion, let us summarize our 
introductory commentaries. We are essentially pointing out
three major problems in the learning and teaching of physics:
1) the 
demand of the physics instructors for effective teaching strategies 
that explain how much time should be
spent on teaching intuitive conceptual reasoning and how much time 
on developing students' quantitative reasoning, and how to teach 
both aspects holistically 
2) the students' 
need for suitable textbooks 
that will help them develop
mathematical abilities reasoning, which are essential 
for enhancing their 
knowledge of conceptual physics,
and 3) a deficiency in the teaching of physics leading to students not being
taught a coherent physics problem-solving strategy that enables them to engage
in both mathematical and conceptual reasoning.

A moment of though about the above summarized
difficulties leads us to postulate the necessity of a systemic 
\cite{Bunge:2000, Bunge:2004} approach
which, from
an operational point of view, could help instructors and
students to achieve a better performance in the process of
teaching and learning physics. 

In a broad sense, a systemic approach in the learning and
teaching of physics could be represented as a framework
involving the ordered
triple composition-environment-structure together with a mechanism or
modus operandi which  integrates the teaching and learning process according
to an approach allowing us to tackle the aforementioned difficulties
from an efficient and unifying point of view 
(for more details refers to Bunge \cite{Bunge:2000, Bunge:2004}).

On the instructor side the need of a systemics approach in the teaching
of physics could be justified by the advantage of using a methodology
which would help them to incorporate both conceptual and mathematical
reasoning systematically in their teaching. In this way, students will
obtain the necessary training in their computational skills while
learning how to use mathematical formulae to obtain the physics
in the equations, even when they can obtain the mathematical solutions of
a problem by
rote procedures. 
In other words,
students could apply ``higher order thinking skills."\cite{Rigden:1987}
via the
mathematical understanding of a physics problem, which in turns involves
meaningful learning which goes beyond the merely application of rote 
procedures.
Moreover, using properly designed quantitative problems
that require
 students to illustrate their conceptual
learning and understanding will reveal much to 
instructors about their students' learning and will provide
invaluable feedback\cite{Reif:1981,ReifScott:1999,YapWong:2007,
DunnBarbanel:2000}, and
such problems can also be a powerful
way to help students to understand the concepts of
physics\cite{Rigden:1987,DunnBarbanel:2000},
a point emphasized 
by the Nobel prize-winning, great physicist
Lev Davidovich Landau on the importance of first mastering  the 
techniques of
working in the field of interest because ``fine points will come
by itself." In Landau's words, ``You must start with mathematics which,
you know, is the foundation of our science.  [...] Bear in mind that
by 'knowledge of mathematics' we
mean not just all kinds of theorems, but a practical ability to
integrate and to solve in quadratures ordinary differential equations,
etc."\cite{Lifshitz:1977}
To further enhance their reasoning skills,
the students would have the opportunity to
increase their intuitive conceptual skills in the physics laboratory,
where conceptual learning is reinforced by 
experience\cite{AufschnaiterAufschnaiter:2007,Hanif:2009}.

On the student side, the need of a systemics approach in the learning
of physics could be 
justified by the usefulness of applying a working methodology which 
could help them to
approach the learning of physics from a interrelate point of view. That
is, that his/her knowledge of mathematics is useful to master ideas
from physics, and
that the use of analogies are important to approach the solution
of physical, mathematical and engineering problems. In short, 
this kind of practical, unified 
problem-solving strategy will help students to pose and approach
any kind of problem, 
and will teach students that physics is the 
primary subject to start
developing these kind of analytical skills. 
In other words, with
such an approach, students would internalize
that it is in physics classes where they can
start to apply what they have learned in their math classes and to
find new non-formal approaches to performing
computations\cite{Yeatts:1992}.  
One could resort to
the anecdote of the
cathedral building\cite{anecdote:2007,anecdote:note} to further illustrate
the need of keeping in mind the interconnectedness of reasoning skill, 
mathematics and physics.
To paraphrase Heron and Meltzer, learning to approach problems in a 
systematic way starts from teaching
and learning the interrelationships among conceptual knowledge,
mathematical
skills and logical reasoning\cite{HeronMeltzer:2005}.
An example of this assertion
is illustrated by an explanation of what happened
to the Millennium Bridge disaster,
. which 
 stated ``Existing theories
of what happened on the bridge's opening day focus on the wobbling of the
bridge but have not addressed the crowd-synchronization dynamics. In
our approach, wobbling and synchrony are inseparable. They emerge together,
as dual aspects of a single instability mechanism, once the crowd reaches a
critical size.''\cite{StrogatzEtAl:2005} The mistakes made in the 
construction 
of the Millennium Bridge help us to understand the need for teaching and
learning based on a systemic approach which recognizes the
interrelatedness of every
aspect of the physical process (physics, mathematics, and engineering
design).  

\section{A systemic structured methodology to solve 
         physics problems}

Earlier work on the importance and necessity of a problem-solving 
strategy can
be found in the work of the great mathematician George P\'olya
\cite{Polya:1945, Polya:1963}, who made emphasis on the relevance of
the systematicity
of a problem-solving
strategy
for productive thinking, discovery and invention. Some of his
views, either provocative or encouraging, about teaching and learning can be 
found spread out
in some PER publications,
like for instance that 
{\em teaching is not a science} (i. e. \cite{Hammer:1996});
on {\em the aim of teaching} (i. e. \cite{Rigden:1987, Redish:2005});
that {\em teaching is an art} (i. e. \cite{Milner:2007}); on
the importance of {\em problem-solving skills} 
(i. e. \cite{Reif:1981, ReifScott:1999, Ehrlich:2002, Heller:1992a}), etc.
For
a
further detailed account of P\'olya's work let's refers to
\cite{Polya:1945, Polya:1963, Schoenfeld:1985, Schoenfeld:1987, 
Schoenfeld:1992,Lederman:2009}.

In {\em How to Solve It}, P\'olya set four  general steps to be followed
as a problem-solving strategy:
\begin{enumerate} 
\item[P1]
\label{P1a}
{\bf Understanding the problem}: some considerations to develop at
this step involves drawing a figure and asking questions like
 What is the unknown? What is the condition?
Is it possible to satisfy the condition? Is the condition sufficient
to determine the unknown? Or is it insufficient? Or redundant? Or 
contradictory? Draw a figure. Introduce suitable notation. 

\item[P2]
\label{P2b}
{\bf Devising a plan}:  some considerations to have in mind in order
to develop this step involves looking at the unknown and trying to
think of a familiar problem having the same or a similar unknown.
Some questions to be ask are like  Have you seen this
before? Or have you seen the same problem in a slightly different form?

\item[P3]
\label{P3c}
{\bf Carrying out the plan}:
Be sure to check each step and make sure that the steps are correct.

\item[P4]
\label{P4d}
{\bf Looking back}: 
 some considerations to develop at this step involves asking questions like
Can the results be checked? Can the results be derived differently? 
Can the result or the method be applied to solve or fully
understand other problems?
\end{enumerate} 

Surprisingly, these steps encompass ``the mental processes and unconscious 
questions experts explore as they themselves approach problem 
solving.''\cite{Lederman:2009} These four step are also the base of
some computational models devised to ``model and explore scientific 
discovery processes.''\cite{Lederman:2009}

Now, even though the aforementioned four steps seems very simply, their
generality seems to be hard for novices to follow them. Thus, in order
to have a problem-solving strategy more affordable to students
we extended the four steps problem-solving strategy into a six steps 
strategy. We made our choice based on empirical observations after
experimenting with a five step strategy reported in \cite{Heller:1992a}.
Justification for a more detailed problem-solving strategy
can be found in the words of Schoenfeld ``First, the strategies are
more complex than their simple descriptions would seem to indicate.
If we want students to use them, we must describe them in detail and 
teach them with the same seriousness that we would teach any other
mathematics.''\cite{Schoenfeld:1980}
Nevertheless, we will further rationalize below the need for explicitly 
including a new
step in our proposed methodology (see item \ref{M5}).
Thus, our six steps proposed strategy is as follows:

\begin{enumerate}
\item
\label{M1}
{\bf Understand the problem}: in this stage students needs to actually be sure
to what the problem is. In addition to making drawings to get a grasp
of the problem formulation, eventually they  might need to
reformulate the problem in their own words, making that that
they are obtaining all the giving information for solving the problem. 
This is a crucial
step in the sense that {\em if one does not know where are we going, 
any route will take us there}.

\item
\label{M2}
{\bf Provide a qualitative description of the problem}: in this stage students 
needs to think and write down the laws, principles or possible formulations
that could help them to solve the problem. For instance students needs 
to consider 
any possible frame work of analysis that could help them to represent
or describe the problem in terms of the principles of physics 
(i.e. Newtons law, 
energy conservation, momentum conservation, theorem of parallel axis for 
computing inertia moment, non-inertial reference system, etc.) If necessary,
the drawings 
of the previous step could be complemented by the corresponding
free-body and/or vector diagram.

\item
\label{M3}
{\bf Plan a solution}: once the student have as many possibilities to
approach the problem, he/she only needs to pick one strategy of solution
and write down the corresponding mathematical formulation of the problem,
avoiding as much as possible to plug numbers in the respective equations.
Also, they need to think whether the problem at hand is similar to a 
previously solved one, and find out whether the information at hand
would be enough to get a solution (i.e. if a set of algebraic 
equations is under or over determined or the number of provided 
boundary conditions are enough 
to solve a differential equation). Eventually, one might need to go back to
the previous step in order to get the a well posed problem, perhaps by 
choosing another strategy. 

\item
\label{M4}
{\bf Carrying out the plan}:
at this stage the student will try to find a solution to the mathematical
formulation of the problem and perhaps they will need to go back
in order to find a easier mathematical formulations of the problem. This
is facilitated is the students have writing down alternatives of solution
as they were suppose to do on item \ref{M2}.

\item
\label{M5}
{\bf Verify the internal consistency and coherence of the used equations}:
at the moment of finding a solution of the involved mathematical equations,
students need to verify whether the equations are consistent with
what they represent and that the equations are dimensionally correct. Though 
this seems to be an unnecessary step, experience
shows that the students too often does not verify the consistency and
coherence of the equations they solve. And this mistake is also
found to be performed by textbook writers, as discussed in a recent 
editorial \cite{Bohren:2009}. After verifying no mistakes or inconsistencies
are found in the mathematical solution of the problem, students could 
then plug numbers in the obtained results to find,
whether required or not, a numerical solution which in turn
could be used in the next step to further evaluate the obtained result. 
In the provided illustrative
example will show how a right answer could be obtained, though
the internal consistency of a used equation is not right\cite{note:2eq1}.    

\item
\label{M6}
{\bf Check and evaluate the obtained solution}:
once a solution have been obtained, its plausibility needs to be evaluated.
Some questions could be asked in this regards
can the results be derived differently? Can the solution be used to 
write down the solution of a less general problem? Can the solution
be used to further understand the qualitative behavior of the problem?
Is it possible to have a division by zero by changing a given parameter? 
Does it makes sense?, and so on.
\end{enumerate}

A first comment on our six  steps problem-solving strategy is that it
provides a unified, systemic way of approaching the solution
of a physical problem encompassing both qualitative (steps \ref{M1}-\ref{M3})
and quantitative (steps \ref{M4}-\ref{M6})
reasoning, and instructors could make as much emphasis as they 
prefer on any of the set of steps, providing the students with a
structured recipe on how to approach in detail the other side of 
the problem's
 solution. 
Second, comparing our six steps problem-solving strategy with P\'olya's four
steps ones, one could see that we have explicitly divided 
P\'olya's step one (P1) into two steps (\ref{M1}-\ref{M2}), and
P\'olya's step three (P3) into two steps (\ref{M4}-\ref{M5}). A
further comment on this problem-solving strategy is that we prefer 
to call the the second
step (\ref{M2}) 
{\em Provide a qualitative description of the problem} rather
than {\em Physics description} as in \cite{Heller:1992a}, because one
share the idea that students tend to think that by providing a qualitative
analysis of a problem they are also providing the solution required
by a physicist, and that the mathematical solution of the problem
is just uninteresting mathematics. Instead, we make emphasis in that
a physical solution of a problem is a combination of both qualitative 
and quantitative reasoning. As 
 stressed by the great physicist Lord Kelvin ``I often say
that when you can
measure something and express it in numbers, you know something about it.
When you can not measure it, when you can not express it in numbers, your
knowledge is of a meager and unsatisfactory kind. It may be the beginning
of knowledge, but you have scarcely in your thoughts advanced to the state
of science, whatever it may be." \cite{Hewitt:1993} Freeman Dyson was
more eloquent ``...mathematics is not just a tool by means of which phenomena
can be calculated; it is the main source of concepts and principles by means of which new theories can be created."\cite{Dyson:1964} 
\section{Illustrative Example}

In the following example we will present an approach on how to
introduce students in the use of our proposed six steps problem-solving
strategy. Though each one of the steps has its importance,
we will provide further evidence of why step
five needs to be taught explicitly. 
It is pertinent to point out that, 
paraphrasing P\'olya's words,
by 
proper training students
could absorb the steps of our problem-solving strategy in such a way
that they could perform the corresponding  operations mentally, naturally,
and vigorously.
The dynamics of teaching is left to the
instructor. In this article we are not pretending to show how
the teaching should be carried out. Innovative teaching strategies
can be found elsewere\cite{YapWong:2007,HoellwarthEtAll:2005,
CrouchMazur:2001, James:2006,CerbinKopp:2006, Heller:1992a, Heller:1992b}.

\begin{figure*}[!htbp]
           \begin{center}
\includegraphics[scale=.4]{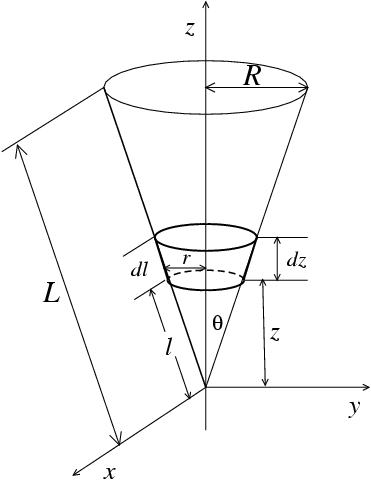}
           \end{center}
\vspace{-0.1in}
 \caption{ a hollow right circular cone with radius R,
lateral length L, and uniform mass M.  The cone's
high is $H=\sqrt{L^2 + R^2}$. The figure also shows
at the lateral distance $l$, measured from the cone's apex
at the origin of the coordinate system, an infinitesimal ring
of lateral length $dl$ and radius $r$. Two useful geometrical relations
among some of the dimensions  shown in
the figure are
$r=(R/L)\,l$ and $r = (R/H)\,z $. }
  \label{conicalfig}
\end{figure*}
\subsection{The problem statement}
  {\bf Problem}: About its central axis, find the moment of inertia of 
a thin hollow right circular cone 
with radius $R$, lateral length $L$, and mass $M$ uniformly distributed on its
surface with density $\sigma$.

\begin{enumerate}
\item
{\bf Understand the problem}: ``It is foolish to answer a question that you
do not understand. $\cdots$ But he should not only understand it, he should
also desire its solution.''\cite{Polya:1945} Following P\'olya's
commentary, before attempting to solve this problem,
students need to have been exposed to a basic theory
on computing moment of inertia ($I$).
Particularly, students need to be familiar with the computation of $I$
for a thin circular ring about its main symmetric axis. To further
understands the geometry of the present problem, students
could,
for
example, be talked about the shape of an empty ice cream cone. 
After some discussion, a drawing 
better than the one shown in figure \ref{conicalfig} could be presented 
on the board. Let's mention that additional ways of presenting each
step in meaningful ways can be found in \cite{Polya:1945,Heller:1992a}.

\item
{\bf Provide a qualitative description of the problem}: In this step one could
further motivate the discussion by associating the computation of $I$
with rotational motion quantities (i.e. kinetic energy,
angular momentum, torque, etc.). One can even motivate the qualitative 
discussion by considering the hollow cone as a first crude approximation 
of a symmetric top or of a cone concrete mixer.
The drawing of figure \ref{conicalfig} could even be made more explicative.

\item
{\bf Plan a solution}: ``We have a plan when we know, or know at least
in outline, which calculations, computations, or constructions we 
have to perform in order to obtain the unknown. $\cdots$ We know,
of course, that it is hard to have a good idea if we have little
knowledge of the subject, and impossible to have it if we have no
knowledge. $\cdots$ Mere remembering is not enough for a good idea,
but we can not have any good idea without recollecting some
pertinent facts.''\cite{Polya:1945} Accordingly, at this stage 
instructors could point out the superposition principle to
solve the problem by slicing the hollow cone in a set of small, infinitesimal,
 rings distributed along the symmetrical axis of the cone. Thus, each 
infinitesimal ring will have in common the same rotational axis about which
the moment of inertia of them is already known $dI = r^2 dm = r^2 \sigma dS$,
where $r$ is the radius of each ring, while $dS$ represents the respective
infinitesimal surface of each ring.

\item
{\bf Carrying out the plan}: To carried out the plan, it won't be a surprise
to choose the wrong  $dS$. In fact, it is not difficult, at first sight,
to choose wrongly (see figure \ref{conicalfig}):
 $dS = 2\pi r dz = 2\pi (R/H) z dz$, which lets to
 $S=\pi R H$, as the hollow cone surface (this result is of course wrong). 
Using this surface element, the momemnt of inertia for the small ring
takes the form
$dI = 2\pi \sigma (H/R) r^3 dr$, which lets to 
$I = 2\pi \sigma (H/R) (R^4/4) = (1/2) (\sigma S) R^2 = M R^2/ 2$, as the
required moment of inertia of the hollow cone (which is the right answer).
It is not difficult to get students performing this sort of
computations and they become uneasy when trying to convince them that
in spite of having found a
right result, it is specious  because it was obtained via a
wrong choice for $dS$. Eventually students might agree on the
incorrectness of their procedure if asked to compute explicitly
the cone's mass. 

\item
{\bf Verify the internal consistency and coherence of the used equations}:
``Check each step. Can you see clearly that the step is correct? Can you
prove that it is correct? $\cdots$ Many mistakes can be avoided if, carrying
out his plan, the student check each step.''\cite{Polya:1945}
 Steeping on our teaching
experience, it is
too easy for students to perform without hesitation the just 
aforementioned 
wrong computations, as presented in the previous step. And it is not 
easy to get students to realize their mistake.
For god shake, they have
computed the right answer!!!: for a hollow thin cone, rotating about
its symmetric axis, $I=M R^2/2$ !!!. In this situation, to make aware
students of their mistake, the easy way is the experiment. 
Instructors could unfold several hollow cones to actually show the
students that the respective surface is $S=\pi R L$, instead of 
the wrongly obtained $S= \pi R H$. Accordingly, we hope to
have provided enough evidence for the need to,
explicitly and repeatedly, mention to students on the need
to check each computational step, including checking for dimensionality
correctness. In this case, the right approach 
is to consider
$dS = 2 \pi r dl =  2 \pi  l (R/L) dl$, which yield
$S= \pi  R L$, the right answer for $S$. This choice for $dS$ lets to
$dI =  2 \pi \sigma (L/R) r^3 dr$, which yields 
 $I =  2 \pi \sigma (L/R) (R^4/4) = (1/2)(\sigma S) R^2 = M R^2/2$,
the right answer.

Considering that it is not hard to find stories on reported wrong 
results due to wrong
or incomplete computations \cite{StrogatzEtAl:2005,Veysey:883}, this
problem could also be used as an example of how computations of a
physical quantity (the surface of a cone shell) can be used to judge
a mathematical result (the wrong value for $S$) that is used in
additional computations yielding a right answer..

\item
{\bf Check and evaluate the obtained solution}:
``Some of the best effects may be lost if the student fails to reexamine and
to reconsider the completed solution.''\cite{Polya:1945} After gaining 
confidence on 
the obtained solution of the problem, it is necessary to spend sometime
in evaluating its plausibility. Examining  the solution of our problem
one could ask:
it is not striking that the rotational inertia for a hollow cone about its
symmetric axis is the same as for a solid disk having the same uniformly
distributed mass $M$ and  radius equal to
the cone's base? 
Does not it a counter example
for the statement
that rotational inertia only depends on how the mass 
is distributed around the axis of rotation? Furthermore, if for some reason
the wrong choice for the $dS$ was not caught in the previous step,
it could be detected if analyzing the case of having a non constant $\sigma$.
A further interpretation of the result can be found at \cite{Bolam:1961}.

\end{enumerate}

\section{Concluding remarks}
This article presents a six step problem-solving
strategy, aiming to approach
three major problems in the learning and teaching of physics:
1) the
demand of the physics instructors for effective teaching strategies
that could help in the
teaching
of intuitive conceptual and quantitative reasoning, and how to teach
both aspects holistically
2) the students'
need for suitable methodology that could help students to fill the textbooks'
gap on
enhancing their
mathematical reasoning abilities, which are essential
for reinforcing students'
knowledge of conceptual physics,
and 3) a deficiency in the teaching of physics leading to students not being
taught a coherent physics problem-solving strategy that enables them to engage
in both mathematical and conceptual reasoning.

Let's finish by recalling
a particular point of view which 
the great mathematician P\'olya stressed very much in his 
writings about the art of teaching and learning,
which, in some sense, can be considered as an ``axiomatic thought''
 about the art of teaching an learning.
He was emphatic on the fact that 
``{\bf {\em for efficient learning, the learner should be interested in 
the material to be learnt and find pleasure in the activity of learning}}.''
In order to reinforce the content of this expression one might
recall the story of the Cathedral's construction workers \cite{anecdote:note}.
In other words, inspiration to learn is without doubt a necessary condition 
in order to
have an efficient and effective teaching and learning environment. This, of 
course, is
by no means a new discovery, and, paraphrasing  
Schoenfeld \cite{Schoenfeld:1977}, some ideas to circumventing
few of the barriers between the dedicated instructor and his/her students'
attitudes in ``learning'' the subject that is being taught 
has been set forward in \cite{Schoenfeld:1977,Ehrlich:2007,Duda:2008}. 
Nevertheless, one should keep in mind that ``we know
from painful experience that a perfectly unambiguous and correct 
exposition can be far from satisfactory and may appear uninspiring, 
tiresome or disappointing, even if the subject-matter presented is
interesting in itself. The most conspicuous blemish of an otherwise acceptable
presentation is the 'deus ex machina'.''\cite{Polya:1949} 
\section*{Acknowledgments}
  We are grateful to
  Dr. Cheryl Pahaham, who kindly provided
useful comments
  on improving this article.

\bibliographystyle{unsrt}
\bibliography{../Article_new_rmf/RojasS_submission_ref}

\begin{thebibliography}{10}
\bibitem{anecdote:note}
The story goes like this. A strange fellow approached three construction
  workers who were part of the team building a cathedral. ``Excuse me, young
  man, what are you doing?'' the strange guy asked one of the workers, who was
  doing his work with some anxiety. ``I am cutting stones,'' the worker
  replied. Realizing that the construction worker had no desire to engage in
  conversation, the strange man repeated his question to another worker. This
  time, this construction worker paused and stated that he was a stonecutter
  from the north who had come to work and who would return home as soon as he
  had earned some money. After thanking the second worker, the strange man
  approached the third construction worker, who was performing his work with
  enthusiasm, patiently and carefully making and remaking was his was doing,
  and again asked his question. This construction worker completely stopped
  what he was doing as if to begin a long discussion with the strange man, and
  answered: ``I have journeyed many miles to be part of the team that is
  constructing this magnificent cathedral. Even though I have spent many months
  away from my family and I miss them dearly, I am aware of how important this
  cathedral will be one day and I know how many people will find spiritual
  peace in this great place.'' This worker continued his story, telling the
  strange man about the planning of the construction of the cathedral, and how
  the construction would be completed, and so on. Relating the anecdote with
  students behavior, the first worker appears like the students who painfully
  are taking a course and only want a passing grade. The second worker is like
  the students who are doing just enough to obtain their passing grade. The
  third worker is like the active-learner students, who have realized that
  their university life is much more than obtaining a passing grade, and that
  every subject is interconnected each to another. Thus, several perspectives
  can be applied to study any subject.
\bibitem{note:2eq1}
In order to further show students the necessity of continuosly verifying the
  consistency of the used equations, one could resort to the naive problem
  involving the wrong proof that $1=2$: $x\times x - x\times x = x^2-x^x
  \rightarrow x(x-x) = (x-x)(x+x) \rightarrow \text{ (after cancelling (x-x) in
  both sides) } x = 2 x \rightarrow 1 =2 $.
\end{thebibliography}
\label{LastPage}
\end{document}